\documentclass{aa}

\usepackage{graphics}

\begin{document}

   \thesaurus{12(05.01.1, 04.03.1, 08.02.3)}

   \title{The combination of ground-based astrometric compilation catalogues
          with the HIPPARCOS Catalogue}

   \subtitle{II. Long-term predictions and short-term predictions}

   \author{R. Wielen
   \and H. Lenhardt
   \and H. Schwan
   \and C. Dettbarn}

   \offprints{R. Wielen (wielen@ari.uni-heidelberg.de)}

   \institute{Astronomisches Rechen-Institut, Moenchhofstrasse 12-14,
   D-69120 Heidelberg, Germany}

   \date{Received 2 August 2000 / Accepted 12 December 2000}

\authorrunning{R. Wielen et al.}

\titlerunning
{The combination of ground-based astrometric compilation catalogues
          with the HIPPARCOS Catalogue. II.}

\maketitle

\begin{abstract}
The combination of ground-based astrometric compilation catalogues, such as the
FK5 or the GC, with the results of the ESA Astrometric Satellite HIPPARCOS
produces for many thousands of stars proper motions which are significantly
more accurate than the proper motions derived from the HIPPARCOS observations
alone. In Paper I (Wielen et al. 1999, A\&A 347, 1046)
we have presented a method of
combination for single stars (SI mode). The present Paper II derives a
combination method which is appropriate for an ensemble of `apparently
single-stars' which contains undetected astrometric binaries. In this case the
quasi-instantaneously measured HIPPARCOS proper motions and positions are
affected by `cosmic errors', caused by the orbital motions of the photo-centers
of the undetected binaries with respect to their center-of-mass. In contrast,
the ground-based data are
`mean values' obtained from a long period of observation. We derive a linear
`long-term prediction' (LTP mode) for epochs far from the HIPPARCOS epoch
$T_{\rm H} \sim 1991.25$, and a linear `short-term prediction' (STP mode) for
epochs close to $T_{\rm H}$. The most accurate prediction for a position at an
arbitrary epoch is provided by a smooth, non-linear transition from the STP
solution to the LTP solution.

We present an example for the application of our method, and we discuss the
error budget of our method for the FK6 (a combination of the FK5 with
HIPPARCOS) and for the combination catalogue
GC+HIP. For the basic fundamental
stars, the accuracy of the FK6 proper motions in the LTP mode is better than
that of the HIPPARCOS proper motions (taking here the cosmic errors into
account) by a factor of more than 4.

\end{abstract}

%Section 1
\section{Introduction}

In Paper I (Wielen et al. 1999b), we have shown that
the combination of the data of the HIPPARCOS astrometric satellite (ESA 1997)
with ground-based results (such as the FK5) is able to provide for many stars
individual proper motions which are significantly more accurate than the
HIPPARCOS proper motions themselves. The method has been already successfully
applied in the construction of the FK6, the Sixth Catalogue of Fundamental
Stars (Part I of the FK6: Wielen et al. 1999d,
Part III of the FK6: Wielen et al. 2000a).

The method of combination presented in Paper I is, however, strictly valid for
single stars only. In the FK6, we call this procedure therefore the
`single-star mode'.

In reality most of the stars are members of binaries or of multiple systems. If
the duplicity of an individual object is already definitely known, either from
ground-based investigations or from HIPPARCOS observations, then the method of
combination has to be changed properly in order to obtain meaningful results.
We call such procedures `special solutions'. Paper III of this series of papers
will discuss the special solutions for visual binaries and other types of
double stars.

%%%%%%%%%%%%%%%%%%%%%%%%%%%%%%%%% Begin of Figure 1 %%%%%%%%%%%%%%%%%%%%%%
% Begin of        %
%                 %
% F i g u r e   1 %
%                 %
%%%%%%%%%%%%%%%%%%%
\begin{figure}[t]
\resizebox{\hsize}{!}{\includegraphics{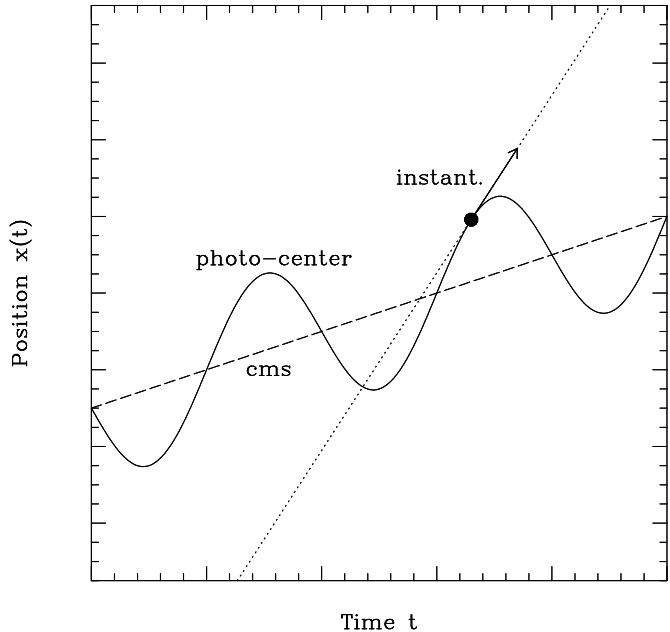}}
\caption[]
{Wavy motion of the photo-center of an astrometric binary (solid
curve) around the linear motion of its center-of-mass (cms; dashed line).
An instantaneous position
and an instantaneous proper motion are indicated (filled dot; arrow).
The linear
prediction based on the instantaneous values is shown as a dotted line.
The long-term averaged 'mean' proper motion is equal to the
motion of the cms.}

\end{figure}
%%%%%%%%%%%%%%%%%%%%%%%%%%%%%%%% End of Figure 1 %%%%%%%%%%%%%%%%%%%%%%%%%%%%%

The present paper (Paper II) describes an appropriate method of the combination
of HIPPARCOS results with ground-based observations for `apparently single
stars'. Even if we have removed from such a sample of apparently single stars
all the objects with known duplicity, there remains beside the truly
single-stars a large number of hitherto undetected astrometric binaries. The
measured photo-center of an unresolved astrometric binary moves on the sky on a
wavy curve, in contrast to the linear motion of single stars (Fig. 1). In such
a case, an `instantaneously' measured proper motion deviates from a `mean'
proper motion, averaged over a long interval of time. The proper motions
provided by HIPPARCOS (ESA 1997) are essentially such instantaneously measured
quantities, since they are derived from positional measurements spread over
about three years only. In contrast, the proper motions given in the FK5
(Fricke et al. 1988, 1991) are long-term averages over up to about 200 years.
We have called the difference between the instantaneous proper-motion and the
mean one the `cosmic error' of the instantaneous proper motion (Wielen 1995a,
b, 1997, Wielen et al. 1997).

In some cases, the individual cosmic error is so large that the duplicity of an
apparently single star is strongly indicated by this fact. We have called such
objects `$\Delta\mu$ binaries' (Wielen et al. 1999a). In most cases, however,
the cosmic error in the HIPPARCOS proper motion of a hitherto undetected
astrometric binary is not {\it individually} significant, but can be shown to
exist only {\it statistically} in a larger sample of apparently single stars.
Nevertheless, an appropriate combination method should not neglect the
statistical consequences of these cosmic errors. Our comparison of HIPPARCOS
proper motions with ground-based results has shown that, at least for brighter
stars, the average cosmic error in a HIPPARCOS proper motion is often larger
than the HIPPARCOS measuring error, typically by a factor of three (Wielen
1995a, b, Wielen et al. 1997, 1998, 1999c).

Wielen (1997, henceforth called Paper WPSA) has developed a coherent scheme of
`statistical astrometry' for handling the effects of cosmic errors in
high-precision astrometry. In the following sections, we shall apply the
concepts of statistical astrometry to the problem of combining the HIPPARCOS
results with ground-based measurements for a sample of apparently single stars.
The main results will be solutions which we call the `long-term prediction mode
(LTP)' and the `short-term prediction mode (STP)'. Our results have already
been applied for the LTP and STP solutions given in Part I and Part III
of the FK6 (Wielen et al. 1999d, 2000a).

It is clear that in principle the most desirable solution for our problem would
be to treat each star individually and fully correctly. This would mean (1) for
truly single stars: to use the `single-star mode (SI)' described in Paper I,
and (2) for binaries: to apply individual orbital corrections, as e.g. done for
Polaris by Wielen et al. (2000b). However, since for `apparently single stars'
the true nature (single or double) of the individual objects is unknown, we
{\it have} to rely on statistical methods in order
to handle such a sample of stars
properly. The treatment of a sample of apparently single stars by our
statistical procedures gives {\it on average} the best astrometric prediction,
and it provides the most realistic error budget for such a sample. In this
sense, our statistical treatment is certainly much more appropriate than to
ignore the binary nature of a large fraction of a sample of apparently single
stars altogether.

In Paper I, we have already pointed out that the older observations carry a
high weight in the combination of ground-based measurements with HIPPARCOS
data, and that therefore the GC (Boss et al. 1937) should be also considered
here because of its high number of rather well-measured stars. We call the
result of the combination of the GC with HIPPARCOS the combination catalogue
GC+HIP.

%Section 2
\section{Basic concepts and equations}

%Section 2.1
\subsection{Concepts}

With respect to the two catalogues which should be combined into a new one, we
follow closely the situation described in Sect. 2 of Paper I. We are using also
as far as possible the nomenclature of that section.

We assume that two astrometric compilation catalogues are available, identified
by the indices 1 (e.g. for the FK5) and 2 (e.g. for HIPPARCOS). For the
combined catalogue (e.g. the FK6), we use the index C, usually now supplemented
by an additional subindex which identifies the special mode of the solution
(e.g. LTP or STP). Each of the two basic catalogues
($i = 1, 2$) provides for
the stars a position $x_i (T_i)$
at a central epoch $T_i$
and a proper motion $\mu_i$
for two angular coordinates
(e.g. right ascension $\alpha_\ast = \alpha
\cos \delta$ and declination $\delta$). The mean measuring errors of $x_i
(T_i)$ and $\mu_i$ are denoted by $\varepsilon_{x, i}$ and $\varepsilon_{\mu,
i}$. Usually one of the catalogues
(e.g. $i = 1$) has first to be reduced to
the astrometrical system of the other catalogue (e.g. the FK5 to the HIPPARCOS
system). In this case, $x_1$ and $\mu_1$ denote the already systematically
corrected quantities, and their mean errors $\varepsilon_{x, 1}$ and
$\varepsilon_{\mu, 1}$ include the uncertainty of the systematic corrections.
For the determination of the systematic differences between two catalogues,
we use methods developed for the construction of the FK5 (Bien et al. 1978).

With respect to the principles of statistical astrometry, we make now the
important additional assumptions that Catalogue 1 gives `mean' quantities for
$x$ and $\mu$, averaged over a long interval of time, while Catalogue 2
provides `instantaneously' measured values of $x$ and $\mu$. If we apply our
scheme to a combination of the FK5 with HIPPARCOS, both assumptions are
fulfilled to a high degree of approximation: The FK5 is based on ground-based
observations spread over about two centuries, while the HIPPARCOS results are
obtained from measurements made during a short interval of time, about three
years only. In the terminology of statistical astrometry, our assumptions mean
that we suppose that the FK5 is free from cosmic errors. The cosmic errors in
the HIPPARCOS positions and proper motions are denoted by $c_x = (\xi
(0))^{1/2}$ and $c_\mu = (\eta (0))^{1/2}$. The correlation functions $\xi
(\Delta t)$, $\eta (\Delta t)$, and $\zeta (\Delta t)$ are explained in WPSA
(especially Sect. 3). Numerical values for $c_x (p)$ and $c_\mu (p)$ as
functions of the parallax $p$ will be provided in Sect. 5.

%Section 2.2
\subsection{Basic equations}

In the combination of two astrometric catalogues of which at least one contains
instantaneously measured data (affected by cosmic errors), the predicted
position $x_{\rm p} (t)$ at an arbitrary epoch $t$ should not be anymore a
linear
function of time. According to the principles of statistical astrometry, the
`best' prediction $x_{\rm p} (t)$ for the true instantaneous position $x (t)$
is
given by the non-linear expression (WPSA, Eq. (82), with a slight change
in nomenclature):

\begin{eqnarray}
x_{\rm p} (t) & = & x_1 (T_1) + \mu_1 \, (t - T_1) \nonumber\\
        & + & \gamma (t) \,
        \big(x_2 (T_2) - x_1 (T_1) -  \mu_1 \, (T_2 - T_1)\big)
                                                                  \nonumber\\
        & + & \beta (t)  \,
        \big(\mu_2 (T_2) - \mu_1\big) \, (t - T_2) \,\, .
\end{eqnarray}                                                            %(1)
The quantities $\gamma (t)$ and $\beta (t)$ are functions of time $t$ to be
determined. For computational reasons, we often use
\begin{equation}
B (t) = \beta (t) (t - T_2)                                               %(2)
\end{equation}
instead of $\beta (t)$.

The functions $\gamma (t)$ and $\beta (t)$ are determined from the condition
that the mean error $\varepsilon _{x, {\rm p}} (t)$ of the predicted positions
$x_{\rm p}
(t)$ should be a minimum for every value of $t$, averaged over the ensemble:
\begin{equation}
\varepsilon^2_{x, {\rm p}} (t) =\,< \big(x_{\rm p} (t) - x (t)\big)^2 >\,= min.
\,\,\, .                                                               %(3)
\end{equation}
The operator $<q>$ means, as in WPSA (1997), the average of the quantity $q$
over the ensemble of similar stars. For an individual star, $<q>$ can be
interpreted as the `expectation value' of the quantity $q$ for this star. In
this sense, we consider the ensemble averages as statistical predictions for
the mean behaviour of a typical individual member of the ensemble.

Inserting Eq. (1) into Eq. (3), and using the scheme of statistical astrometry,
we obtain for $\varepsilon^2_{x, {\rm p}} (t)$:
\begin{eqnarray}
\lefteqn{
\varepsilon^2_{x, {\rm p}} (t)
= \, < \Big( \big (x_{\rm p} (t) - x_{\rm m} (t) \big)
- \big( x (t) - x_{\rm m} (t) \big) \Big)^2 > }           \nonumber\\
& & = \, < \Big( \big( x_1 (T_1) - x_{\rm m} (T_1) \big)
(1 - \gamma (t))    \nonumber\\
& & \hspace{0.70 cm}
+\,( \mu_1 - \mu_{\rm m}) \big( t - T_1 - \gamma (t)(T_2 - T_1)
- \beta (t) (t - T_2)\big) \nonumber\\
& & \hspace{0.70 cm} +\, \gamma (t)
\big((x_2 (T_2) - x (T_2)) + (x (T_2) - x_{\rm m}
(T_2))\big)                                             \nonumber\\
& & \hspace{0.70 cm} +\, \beta (t) (t - T_2)
\big((\mu_2 (T_2) - \mu (T_2)) + (\mu (T_2)
- \mu_{\rm m}) \big) \nonumber\\
& & \hspace{0.70 cm} -\,(x (t) - x_{\rm m} (t))\Big)^2 >         \nonumber\\
& & = \big( 1 + (\gamma (t))^2 \big) \, \xi (0)
- 2\,\gamma (t) \,\xi (t - T_2)     \nonumber\\
& & \hspace{0.70 cm}
+\,\big( \beta (t) (t - T_2) \big)^2 \eta (0) - 2\,\beta (t) (t - T_2)
\zeta (t - T_2)                                                    \nonumber\\
& & \hspace{0.70 cm} +\,\varepsilon^2_{x, 1} \big( 1 - \gamma (t) \big)^2 +
\varepsilon^2_{x, 2} \big( \gamma (t) \big)^2                      \nonumber\\
& & \hspace{0.70 cm} +\,\varepsilon^2_{\mu, 1} \big( t - T_1 - \gamma (t)
(T_2 - T_1) - \beta (t) (t - T_2) \big)^2                          \nonumber\\
& & \hspace{0.70 cm} +\,\varepsilon^2_{\mu, 2} (\beta (t) (t - T_2))^2 \,\,\, .
\end{eqnarray}                                                       %(4)
The quantities $x_{\rm m} (t)$ and $\mu_{\rm m} = \dot{x}_{\rm m}$
are the true mean position
and the true mean proper motion of the star
(in the sense of statistical astrometry).
$x_{\rm m}$ is a linear function of $t$, e.g.
\begin{equation}
x_{\rm m} (t) = x_{\rm m} (T_1) + \mu_{\rm m} \, (t - T_1) \,\, .                         %(5)
\end{equation}
In deriving Eq. (4) we have used the following relations:
According to our assumption that Catalogue 1 provides mean quantities, $x_1
(T_1)$ and $\mu_1$ differ from $x_{\rm m} (T_1)$ and $\mu_{\rm m}$
by their measuring errors only:
\begin{eqnarray}
< (x_1 (T_1) - x_{\rm m} (T_1))^2 > & = & \varepsilon^2_{x, 1} \,\, , \\ %(6)
< (\mu_1 - \mu_{\rm m})^2 >         & = & \varepsilon^2_{\mu, 1} \,\, .  %(7)
\end{eqnarray}
The quantities $x (t)$ and $\mu (t)$ are the true instantaneous position and
proper motion of the star at epoch $t$. Since Catalogue 2 is assumed to provide
instantaneous quantities, $x_2 (T_2)$ and $\mu_2 (T_2)$ differ from $x (T_2)$
and $\mu (T_2)$ also by their measuring errors only:
\begin{eqnarray}
< (x_2 (T_2) - x (T_2))^2 > & = & \varepsilon^2_{x, 2} \,\, ,        \\  %(8)
< (\mu_2 (T_2) - \mu (T_2))^2 > & = & \varepsilon^2_{\mu, 2} \,\, .      %(9)
\end{eqnarray}
The correlation functions $\xi (\Delta t), \, \eta (\Delta t)$, and $\zeta
(\Delta t)$ (see WPSA) occur because of:
\begin{equation}
< (x (T_2) - x_{\rm m} (T_2))^2 >\,=\,< (x (t) - x_{\rm m} (t))^2 >\,
= \xi (0) \, ,                                                           %(10)
\end{equation}
\begin{equation}
< (x (T_2) - x_{\rm m} (T_2)) (x (t) - x_{\rm m} (t)) >\,
= \xi (t - T_2) \,\, ,                                                  %(11)
\end{equation}
\begin{equation}
< (\mu (T_2) - \mu_{\rm m})^2 >\,= \eta (0) \,\, ,                    %(12)
\end{equation}
\begin{equation}
< (x (t) -  x_{\rm m} (t)) (\mu (T_2) - \mu_{\rm m}) >\,
= \zeta (t - T_2) \,\, .                                                 %(13)
\end{equation}
We have further to remember
\begin{equation}
< (x (T_2) - x_{\rm m} (T_2)) (\mu (T_2) - \mu_{\rm m}) >\,
= \zeta (0) = 0 \,\, .                                                   %(14)
\end{equation}
The products of the other terms occuring in the second line of Eq. (4) all
vanish, because the measuring errors are not correlated with the cosmic errors,
for example
\begin{equation}
< (x_1 (T_1) - x_{\rm m} (T_1)) (x (t) - x_{\rm m} (t)) > \,= 0 \,\, ,               %(15)
\end{equation}
and because the measuring errors of $x_i (T_i)$ and $\mu_i (T_i)$ are not
correlated at the central epoch $T_i$ (by definition of the central epoch), for
example:
\begin{equation}
< (x_1 (T_1) - x_{\rm m} (T_1)) (\mu_1 - \mu_{\rm m}) > \, = 0 \,\, ,                %(16)
\end{equation}
and finally because the measuring errors of $x_1$ and $\mu_1$ are not
correlated with those of $x_2$ and $\mu_2$, for example
\begin{equation}
< (x_1 (T_1) - x_{\rm m} (T_1)) (\mu_2 (T_2) - \mu (T_2)) > \, = 0 \,\, .       %(17)
\end{equation}

The conditions for the minimum of $\varepsilon^2_{x, {\rm p}} (t)$ with respect
to $\gamma$ and $\beta$ are
\begin{equation}
\frac{\partial\varepsilon^2_{x, {\rm p}}}{\partial\gamma} = 0 \,\, ,   %(18)
\end{equation}
and
\begin{equation}
\frac{\partial\varepsilon^2_{x, {\rm p}}}{\partial\beta} = 0 \,\, .    %(19)
\end{equation}
Carrying out these procedures and using the final part of Eq. (4) for
$\varepsilon^2_{x, {\rm p}} (t)$, we obtain the following two equations of
condition
for the functions $\gamma (t)$ and $\beta (t)$ which give the `best' values of
$\varepsilon_{x, {\rm p}} (t)$:
\begin{eqnarray}
& & \gamma (t) (\xi (0) + \varepsilon^2_{x, 1} + \varepsilon^2_{x, 2} +
    \varepsilon^2_{\mu, 1} (T_2 - T_1)^2)                        \nonumber\\
\lefteqn{
+ \, \beta (t) (t - T_2) \varepsilon^2_{\mu, 1} (T_2 - T_1)
}                                                                 \nonumber\\
& & = \xi (t - T_2) + \varepsilon^2_{x, 1} + \varepsilon^2_{\mu, 1} (t - T_1)
      (T_2 - T_1) \,\, ,                                      \\[2ex]   %(20)
& & \gamma (t) \varepsilon^2_{\mu, 1} (T_2 - T_1)               \nonumber\\
\lefteqn{
+ \, \beta (t) (t - T_2) (\eta (0) + \varepsilon^2_{\mu, 1} +
\varepsilon^2_{\mu, 2})
}                                                                 \nonumber\\
& & = \zeta (t - T_2) + \varepsilon^2_{\mu, 1} (t - T_1) \,\, .          %(21)
\end{eqnarray}
The Eqs. (20) and (21) correspond to the Eqs. (86) and (87) in WPSA.
Solving the Eqs. (20) and (21) for the unknowns $\gamma (t)$ and $\beta (t)$ we
obtain:
\begin{eqnarray}
& & \gamma (t) = \Big((\xi (t - T_2) + \varepsilon^2_{x, 1}) (\eta (0) +
    \varepsilon^2_{\mu, 1} + \varepsilon^2_{\mu, 2})              \nonumber\\
\lefteqn{
+ \, \varepsilon^2_{\mu, 1} (T_2 - T_1) ((\eta (0) + \varepsilon^2_{\mu, 2})
  (t - T_1) - \zeta (t - T_2))\Big) / N\!\!,
}                                                                      %(22)
\end{eqnarray}
\begin{eqnarray}
B (t) & = & \beta (t) (t - T_2)                                  \nonumber\\
&       = & \Big(\zeta (t - T_2) \big( \xi (0) + \varepsilon^2_{x, 1} +
      \varepsilon^2_{x, 2} + \varepsilon^2_{\mu, 1} (T_2 - T_1)^2 \big)
                                                                  \nonumber\\
& & + \, \varepsilon^2_{\mu, 1} (t - T_1) \big( \xi (0) + \varepsilon^2_{x, 1}
      + \varepsilon^2_{x, 2} \big)                                \nonumber\\
& & -  \, \varepsilon^2_{\mu, 1} (T_2 - T_1) \big( \xi (t - T_2) +
      \varepsilon^2_{x, 1} \big) \Big) / N \,\, ,                     %(23)
\end{eqnarray}
with the auxiliary quantity $N$:
\begin{eqnarray}
\lefteqn{
N = (\xi (0) + \varepsilon^2_{x, 1} + \varepsilon^2_{x, 2}) (\eta (0) +
    \varepsilon^2_{\mu, 1} + \varepsilon^2_{\mu, 2})
}                                                                  \nonumber\\
& & \hspace{0.50 cm} + \, \varepsilon^2_{\mu, 1} (T_2 - T_1)^2 (\eta (0) +
      \varepsilon^2_{\mu, 2}) \,\, .                                     %(24)
\end{eqnarray}

For deriving the functions $\gamma (t)$ and $\beta (t)$, we have up to now
implicitely assumed that all the stars in the ensemble have the same measuring
errors. This is, of course, not strictly true in reality. However, the ensemble
averages are actually neccessary for handling the cosmic errors only, but not
for treating the measuring errors. Hence we shall use all the equations derived
above by inserting the individual measuring errors if we treat individual stars
of the ensemble. Formally we may imagine to handle subsamples of stars in which
the stars have the overall behaviour with respect to the cosmic errors, but in
which the common measuring errors are equal to those of the individual star
under consideration. It is a more severe problem that we do not make use of our
knowledge of how large the individual cosmic errors in $x$ and $\mu$ are for a
given individual object (e.g. $\mu_2 - \mu_1$). As discussed in WPSA,
this would require `conditioned' correlation functions. Since this
information is presently not available, we are treating here the consequences
of the cosmic errors on the level of ensemble averages only.

Inserting these results for $\gamma (t)$ and $\beta (t)$ from Eqs. (22) and
(23) into Eqs. (1) and (4), we derive the prediction $x_{\rm p} (t)$ for the
instantaneous position $x (t)$ of the star and the mean error
$\varepsilon_{x, {\rm p}} (t)$ of this prediction. The prediction $x_{\rm p}
(t)$ is a non-linear
function of $t$, because the correlation functions $\xi (t - T_2)$
and $\zeta (t - T_2)$, which occur in the formulae for $\gamma (t)$ and $\beta
(t)$, are non-linear functions. A typical run of $x_{\rm p}(t)$ is shown in
Fig. 2.

In order to illustrate the properties of our prediction $x_{\rm p} (t)$, we
consider
in the following Subsections 2.3 and 2.4 two limiting cases in which we either
neglect the measuring errors or the cosmic errors.

%Section 2.3
\subsection{Measuring errors neglected}

If we neglect all the measuring errors and set $\varepsilon_{x, 1} (T_1) =
\varepsilon_{x, 2} (T_2) = \varepsilon_{\mu, 1} = \varepsilon_{\mu, 2} = 0$,
then we obtain for $\gamma (t)$ and $\beta (t)$
\begin{eqnarray}
\gamma_{\rm nme} (t) & = & \xi (t - T_2) / \xi (0) \,\, ,      \\     %(25)
B_{\rm nme} (t)      & = & \beta_{\rm nme} (t) (t - T_2) = \zeta (t - T_2) /
\eta (0) \,\, .                                                        %(26)
\end{eqnarray}
These results were already derived and discussed in WPSA (Sect. 4.2.4
and Fig. 10). The corresponding prediction $x_{\rm p, nme} (t)$, shown in
Fig. 2,
passes through the point $x_2 (T_2)$ with the slope $\mu_2 (T_2)$. Therefore,
the measured instantaneous position and proper motion at epoch $T_2$ are
exactly reproduced by $x_{\rm p, nme} (t)$. For $t \rightarrow \pm \, \infty$,
the prediction $x_{\rm p, nme} (t)$ approaches asymptotically the mean position
$x_{\rm m} (t) = x_1 (t) = x_1 (T_1) + \mu_1 \, (t - T_1)$ of the star. The
uncertainty
$\varepsilon_{x, \rm{p, nme}} (t)$ of the prediction is given by
\begin{eqnarray}
\varepsilon^2_{x, \rm{p, nme}} (t)
= \frac{ (\xi (0))^2 - (\xi (t - T_2))^2 }{ \xi (0) }
- \frac{ (\zeta (t - T_2))^2 }{ \eta (0) } \, .                 %(27)
\end{eqnarray}
We see that $\varepsilon_{x, \rm{p, nme}}$ is zero at the epoch $T_2$ and
approaches
$c_x = (\xi (0))^{1/2}$ for $t \rightarrow \pm \, \infty$, since in this limit
$x_{\rm p} (t)$ is equal to $x_{\rm m} (t)$, and $x_{\rm m} (t)$
differs from the instantaneous
position $x (t)$ by the cosmic error $c_x$ on average.

%Section 2.4
\subsection{Cosmic errors neglected}

If we neglect the cosmic errors and set $\xi = \eta = \zeta = 0$, then we
obtain
\begin{eqnarray}
\lefteqn{
\gamma_{\rm{nce}} (t)  =
\Big(\varepsilon^2_{x, 1} (\varepsilon^2_{\mu, 1} +
\varepsilon^2_{\mu, 2})   }                                  \nonumber\\
& & + \, \varepsilon^2_{\mu, 1} \, \varepsilon^2_{\mu, 2} (T_2 - T_1) (t -
T_1)\Big) / N_{\rm{nce}} \,\, ,                                \\[2ex] %(28)
\lefteqn{
B_{\rm{nce}} (t)  =
\beta_{\rm{nce}} (t) (t - T_2) = \Big( - \varepsilon^2_{x, 1}
\varepsilon^2_{\mu, 1} (T_2 - T_1)    }                        \nonumber\\
& & + \, (\varepsilon^2_{x, 1} + \varepsilon^2_{x, 2}) \varepsilon^2_{\mu, 1}
(t - T_1)\Big) / N_{\rm{nce}} \,\, ,                               %(29)
\end{eqnarray}
with
\begin{equation}
N_{\rm{nce}} =
(\varepsilon^2_{x, 1} + \varepsilon^2_{x, 2}) (\varepsilon^2_{\mu, 1}
+ \varepsilon^2_{\mu, 2}) + \varepsilon^2_{\mu, 1} \, \varepsilon^2_{\mu, 2}
(T_2 - T_1)^2 \, .                                                    %(30)
\end{equation}

Equation (29) illuminates one of the advantages of introducing $B (t) = \beta
(t) (t - T_2)$ as a substitute for $\beta (t)$. While $B_{\rm{nce}} (T_2)$,
and in general $B (T_2)$ for non-zero measuring errors, remains finite,
the quantity $\beta_{\rm{nce}}$, and in general $\beta$, tends towards
infinity for $t \rightarrow T_2$. Only in some degenerated cases, such as
$\varepsilon_{\mu, 1} = 0$ or $\varepsilon_{x, 2} = 0$ or $\varepsilon_{x, 1}
\rightarrow \, \infty$, the quantity $\beta_{\rm{nce}}$, and in general
$\beta$, remains finite for $t \rightarrow T_2$.

If we insert $\gamma_{\rm{nce}} (t)$ and
$B_{\rm{nce}} (t)$ into Eq. (1), the
corresponding prediction $x_{\rm{p, nce}} (t)$ for the position of the star at
an epoch $t$ is now a strictly linear function of
$t$, since $\gamma_{\rm{nce}} (t)$
and $B_{\rm{nce}} (t)$ are linear in $t$.

In order to facilitate the understanding of the behaviour of the prediction
$x_{\rm{p, nce}} (t)$, we rewrite $x_{\rm{p, nce}}$ by using the auxiliary
quantities
$T_{\rm{C, nce}}, \, x_{\rm{C, nce}}, \, \mu_{\rm{C, nce}}$. If we insert Eqs.
(28) and
(29) into Eq. (1), we obtain after some lengthy algebra:
\begin{eqnarray}
x_{\rm{p, nce}} (t) & = & x_{\rm{C, nce}} (T_{\rm{C, nce}}) + \mu_{\rm{C, nce}}
\, (t - T_{\rm{C, nce}}) \,\, ,                                           %(31)
\end{eqnarray}
with the auxiliary quantities
\begin{eqnarray}
T_{\rm{C, nce}} & = & \frac{w_{x, 1} T_1 + w_{x, 2} T_2}
{w_{x, 1} + w_{x, 2}} \,\, ,                                      \\[1ex] %(32)
x_{\rm{C, nce}} (T_{\rm{C, nce}}) & = & \frac{w_{x, 1} x_1 (T_1) + w_{x, 2} x_2
(T_2)} {w_{x, 1} + w_{x, 2}} \,\, ,                              \\[1ex] %(33)
\mu_{\rm{C, nce}} & = & \frac{w_{\mu, 1} \mu_1 + w_{\mu, 2} \mu_2 + w_{\mu, 0}
\mu_0} {w_{\mu, 1} + w_{\mu, 2} + w_{\mu, 0}} \,\, ,             \\[1ex] %(34)
\mu_0 & = & (x_2 (T_2) - x_1 (T_1)) / (T_2 - T_1) \,\, .                 % (35)
\end{eqnarray}
The weights $w$ are given by
\begin{eqnarray}
w_{x, 1} & = & \frac{1}{\varepsilon^2_{x, 1}} \,\, ,             \\[1ex] %(36)
w_{x, 2} & = & \frac{1}{\varepsilon^2_{x, 2}} \,\, ,             \\[1ex] %(37)
w_{\mu, 1} & = & \frac{1}{\varepsilon^2_{\mu, 1}} \,\, ,         \\[1ex] %(38)
w_{\mu, 2} & = & \frac{1}{\varepsilon^2_{\mu, 2}} \,\, ,       \\[0.5ex] %(39)
w_{\mu, 0} & = & \frac{1}{\varepsilon^2_{\mu, 0}}
= \frac{(T_2 - T_1)^2}{\varepsilon^2_{x, 1} + \varepsilon^2_{x, 2}} \,\,. %(40)
\end{eqnarray}
Inserting $\gamma_{\rm{nce}} (t)$ and $B_{\rm{nce}} (t)$
into Eq. (4), we obtain the mean
error $\varepsilon_{x, \rm{p, nce}} (t)$ of $x_{\rm{p, nce}}$.
Using the form of $x_{\rm{p, nce}}$
as given in Eq.
(31) and the auxiliary quantities with the index C, we find
\begin{equation}
\varepsilon^2_{x, \rm{p, nce}} (t) = \varepsilon^2_{x, \rm{C, nce}} (T_{\rm{C,
nce}}) +
\varepsilon^2_{\mu, \rm{C, nce}} \, (t - T_{\rm{C, nce}})^2 \,\, ,     %(41)
\end{equation}
with
\begin{equation}
\varepsilon^2_{x, \rm{C, nce}} (T_{\rm{C, nce}}) = \frac{1}{w_{x, 1} + w_{x,
2}} \,\, ,
\end{equation}                                                           %(42)
\begin{equation}
\varepsilon^2_{\mu, \rm{C, nce}} = \frac{1}{w_{\mu, 1} + w_{\mu, 2} + w_{\mu,
0}} \,\, .                                                           %(43)
\end{equation}

A comparison of the Eqs. (31)--(42) with the analytic version of the
single-star-mode solution of Paper I (Sect. 2, especially Eqs. (19), (23),
(30)--(32)) shows that the prediction $x_{\rm{p, nce}} (t)$ and its mean error
$\varepsilon_{x, \rm{p, nce}} (t)$ are strictly identical with the
single-star-mode solution $x_{\rm{SI}} (t)$ and its mean error $\varepsilon_{x,
\rm{SI}} (t)$. This result is
very pleasing, since it proves the internal consistency of our scheme: In
the limit of vanishing cosmic errors, the prediction $x_{\rm p} (t)$ according
to Eq. (1) is asymptotically approaching the single-star mode solution
$x_{\rm{SI}} (t)$.
This result is not apriori self-evident, since our definitions for the
`best' solution for predicting $x (t)$ differ at least formally in Paper I and
in this Paper II (i.e., Eqs. (5) and (35) of Paper I versus Eq. (3) of Paper
II).

%Section 2.5
\subsection{Motivation for introducing the long-term and short-term prediction}

The general solution of our combination problem is given in Sect. 2.2. for all
epochs $t$. The solution $x_{\rm p} (t)$ is a non-linear function of $t$, and
requires the knowledge of the correlation functions $\xi (\Delta t)$ and
$\zeta (\Delta t)$ as functions of the epoch difference $\Delta t$. At present,
we do not have a well-established
knowledge about the run of $\xi (\Delta t)$ and
$\zeta (\Delta t)$. Only the cosmic errors $c_\mu = (\eta (0))^{1/2}$ and
$c_x = (\xi (0))^{1/2}$ can be empirically determined from the comparison of
the FK5 with HIPPARCOS, assuming that the FK5 is giving `mean' quantities.

Even if we would know the run of $\xi (\Delta t)$ and $\zeta (\Delta t)$ as a
function of $\Delta t$, the non-linearity of $x_{\rm p} (t)$ would demand
a table of
$x_{\rm p} (t)$ for a sequence of epochs, e.g. for each year, if
the user should
not have the burden to do the full calculation himself by running a
program.

We propose the following solution: The general solution for $x_{\rm p} (t)$
allows rather easily to obtain two limiting solutions for $\Delta t = t - T_2
\rightarrow \pm \, \infty$ and for $\Delta t \rightarrow 0$. We call the
solution for $\Delta t \rightarrow \pm \, \infty$ the `long-term prediction
(LTP)', and the solution for $\Delta t \rightarrow 0$ the `short-term
prediction (STP)' around the epoch $T_2$.

Both the LTP and STP solutions are linear in $t$. They can be therefore given
in the usual astrometric style, i.e. as a position at a central epoch and a
proper motion, together with their mean errors. The details on the LTP and STP
solutions $x_{\rm{LTP}} (t)$ and $x_{\rm{STP}} (t)$ are given in the Sects. 3
and 4.

The general solution $x_{\rm p} (t)$ is a smooth transition from short-term
prediction $x_{\rm{STP}} (t)$ (for epochs around $T_2$) to the long-term
prediction $x_{\rm{LTP}} (t)$ for epochs $t$ far away from $T_2$. In Sect. 5,
we will discuss a convenient (but only approximately valid) method to carry
out this transition, if we know the run of $\zeta (\Delta t)$.
This gives at least a rough indication on the process of transition as a
function of the epoch difference $\Delta t$, even if $\zeta (\Delta t)$ is
not well-established. If, in the future, $\zeta (\Delta t)$ should be better
determined, then our method would allow rather conveniently the (approximate)
determination of $x_{\rm p} (t)$ also for epochs inbetween the validity ranges
of $x_{\rm{STP}} (t)$ and $x_{\rm{LTP}} (t)$.

%Section 3
\section{Long-term prediction (LTP)}

We consider in this section the limiting case of the general solution
$x_{\rm p} (t)$ for
$|t - T_2| \rightarrow \pm \, \infty$. This `long-term prediction'
$x_{\rm{LTP}} (t)$ is valid for epochs not too close to the epoch $T_2$ of the
instantaneous Catalogue 2 (i.e. here the HIPPARCOS Catalogue with $T_2 \sim
1991.25$).

We assume that the epoch difference $\Delta t = t - T_2$ is so large that the
correlation functions $\xi (\Delta t)$ and $\zeta (\Delta t)$ both vanish.
Setting $\xi (t - T_2) = 0$ and $\zeta (t - T_2) = 0$, we obtain from the
general Eqs. (22) and (23) for the LTP solution
\begin{eqnarray}
\gamma_{\rm{LTP}} (t) & = & \Big(\varepsilon^2_{x, 1}
\left(\varepsilon^2_{\mu,
1} + \left[\varepsilon^2_{\mu, 2} + \eta (0)\right]\right)        \nonumber\\
& &  \hspace{-0.40 cm}
+ \varepsilon^2_{\mu, 1} \left[\varepsilon^2_{\mu, 2} +
\eta (0)\right] \left(T_2 - T_1\right) \left(t - T_1\right)\Big) / N \,\,  ,
                                                                \\[1ex] %(44)
B_{\rm{LTP}} (t) & = & \beta_{\rm{LTP}} (t) \left(t - T_2\right)  \nonumber\\
            & = & \Big(- \varepsilon^2_{x, 1} \varepsilon^2_{\mu, 1}
\left(T_2 - T_1\right)                                            \nonumber\\
& &  \hspace{-0.40 cm}
+ \left(\varepsilon^2_{x, 1}
+ \left[\varepsilon^2_{x, 2}
+ \xi (0)\right]\right)
\varepsilon^2_{\mu, 1} \left(t - T_1\right)\Big) / N \,\, ,        %(45)
\end{eqnarray}
where $N$ is still given by Eq. (24). If we compare the Eqs. (43), (44), and
(24) for $\gamma_{\rm{LTP}}$, $B_{\rm{LTP}}$,
and $N = N_{\rm{LTP}}$ with the corresponding
Eqs. (28)--(30) for $\gamma_{\rm{nce}}$, $B_{\rm{nce}}$,
and $N_{\rm{nce}}$, we find that they are
identical if we replace in the equations for the nce solution the quantity
$\varepsilon^2_{x, 2}$ by
\begin{equation}
\varepsilon^2_{x, 2, \rm{LTP}} = \varepsilon^2_{x, 2} + \xi (0)
= \varepsilon^2_{x, 2} + c^2_x \,\, ,                                   %(46)
\end{equation}
and
$\varepsilon^2_{\mu, 2}$ by
\begin{equation}
\varepsilon^2_{\mu, 2, \rm{LTP}} = \varepsilon^2_{\mu, 2} + \eta (0)
= \varepsilon^2_{\mu, 2} + c^2_\mu \,\, .                              %(47)
\end{equation}
This is very plausible, since the instantaneously measured quantities $x_2
(T_2)$ and $\mu_2 (T_2)$ are affected by the cosmic errors $c_x$ and $c_\mu$.
If we add these cosmic errors to the corresponding measuring errors
$\varepsilon_{x, 2}$ and $\varepsilon_{\mu, 2}$, we obtain `apparent' measuring
errors $\varepsilon_{x, 2, \rm{LTP}}$ and $\varepsilon_{\mu, 2, \rm{LTP}}$.
Since the
cosmic errors are not correlated with the measuring errors, the summation has
to be done quadratically.

Using this finding we obtain for the long-term prediction
\begin{eqnarray}
\hspace*{0.7cm} x_{\rm{LTP}} (t) & = & x_{\rm{LTP}} (T_{\rm{LTP}}) +
\mu_{\rm{LTP}} \,\, (t - T_{\rm{LTP}}) \,\, ,
%(48)
\end{eqnarray}
with\\[-3ex]
\begin{eqnarray}
T_{\rm{LTP}}   & = & \frac{w_{x, 1} \, T_1 + w_{x, 2, \rm{LTP}} \, T_2}
{w_{x, 1} + w_{x, 2, \rm{LTP}}} \,\, ,                      \\[0.5ex]  %(49)
x_{\rm{LTP}} (T_{\rm{LTP}})
& = & \frac{w_{x, 1}\,  x_1 (T_1) + w_{x, 2, \rm{LTP}} \, x_2 (T_2)}
{w_{x, 1} + w_{x, 2, \rm{LTP}}} \,\, ,                  \\[0.5ex] %(50)
\mu_{\rm{LTP}}   & = & \frac{w_{\mu, 1} \, \mu_1
+ w_{\mu, 2, \rm{LTP}} \, \mu_2 +
w_{\mu, 0, \rm{LTP}} \, \mu_0}
{w_{\mu, 1} + w_{\mu, 2, \rm{LTP}} + w_{\mu, 0, \rm{LTP}}} \, .         %(51)
\end{eqnarray}
The weights $w$ are given by Eqs. (36) and (38), and by
\begin{eqnarray}
w_{x, 2, \rm{LTP}} & = & \frac{1}
{\varepsilon^2_{x, 2} + c^2_x} \,\, ,                       \\[0.5ex] %(52)
w_{\mu, 2, \rm{LTP}} & = & \frac{1}
{\varepsilon^2_{\mu, 2} + c^2_\mu} \,\, ,                       \\[0.5ex] %(53)
w_{\mu, 0, \rm{LTP}} & = & \frac{(T_2 - T_1)^2}
{\varepsilon^2_{x, 1} + \varepsilon^2_{x, 2} + c^2_x} \,\, .              %(54)
\end{eqnarray}

The long-term prediction $x_{\rm{LTP}} (t)$ has in fact two conceptionally
different
properties: (1) As described above, it is the limit of the general prediction
$x_{\rm p} (t)$ for the true {\it instantaneous} position $x (t)$ for $t
\rightarrow
\pm \, \infty$. (2) On the other hand, $x_{\rm{LTP}} (t)$ is for all epochs $t$
the best prediction
for the {\it mean} position $x_{\rm m} (t)$ of the object. This means
especially that $\mu_{\rm{LTP}}$ is the best estimate of the center-of-mass
velocity of the object.

The two different concepts produce two different error estimates
$\varepsilon_{x, \rm{LTP}} (t)$ for $x_{\rm{LTP}} (t)$. If we consider
$x_{\rm{LTP}}$ as
the prediction for the {\it mean} position $x_{\rm m} (t)$ then the mean error
$\varepsilon_{x, \rm{LTP, m}} (t)$ is given by:
\begin{equation}
\varepsilon^2_{x, \rm{LTP, m}} (t)  =  \varepsilon^2_{x, \rm{LTP, m}}
(T_{\rm{LTP}}) +
\varepsilon^2_{\mu, \rm{LTP, m}} \, (t - T_{\rm{LTP}})^2 \, ,           %(55)
\end{equation}
with
\begin{equation}
\varepsilon^2_{x, \rm{LTP, m}} (T_{\rm{LTP}})  =  \frac{1}
{w^2_{x, 1} + w^2_{x, 2, \rm{LTP}}} \,\, ,                              %(56)
\end{equation}
\begin{equation}
\varepsilon^2_{\mu, \rm{LTP, m}}  =  \frac{1}
{w^2_{\mu, 1} + w^2_{\mu, 2, \rm{LTP}} + w^2_{\mu, 0, \rm{LTP}}} \,\, . %(57)
\end{equation}
If we consider $x_{\rm{LTP}}$ as the prediction for the {\it instantaneous}
position
$x (t)$ for large values of $|\Delta t| = |t - T_2|$, then the uncertainty
$\varepsilon_{x, \rm{LTP, ins}} (t)$ of $x_{\rm{LTP}} (t)$ is given by
\begin{equation}
\varepsilon^2_{x, \rm{LTP, ins}} (t) = \varepsilon^2_{x, \rm{LTP, m}} (t) +
c^2_x \,\, .                                                           %(58)
\end{equation}
Similarly, the mean error $\varepsilon_{\mu, \rm{LTP, ins}}$
of the predicted
{\it instantaneous} proper motion for large $\Delta t$ is given by
\begin{equation}
\varepsilon^2_{\mu, \rm{LTP, ins}} = \varepsilon^2_{\mu, \rm{LTP, m}}
+ c^2_\mu \,\, .                                                        %(59)
\end{equation}

The equations above describe what we have called the `analytic' approach in
Paper I. For the `numerical' approach, we can take over for the LTP solution
the formulae given in Paper I for the single-star mode (SI) with the following
changes: (1) We should not redetermine the parallax $p$ of the star. The
determination of $p$ by HIPPARCOS requires instantaneous values of $x$ and
$\mu$, not the `mean' LTP values. Formally, we set all correlations between $p$
and the other quantities (position, proper motion) equal to zero. (2) In the
variance-covariance matrix {\bf D} we replace in the diagonal line the
HIPPARCOS values of $D_{\rm{H}, 11} = \varepsilon^2_{\alpha\ast, \rm{H}}
(T_{\rm H})$ by
\begin{equation}
D_{\rm{H, 11, LTP}} = \varepsilon^2_{\alpha\ast, \rm{H}} (T_{\rm H}) + c^2_x
\,\, ,                                                                   %(60)
\end{equation}
and $\varepsilon^2_{\delta, \rm{H}}$ by $\varepsilon^2_{\delta, \rm{H}} +
c^2_x$,
$\varepsilon^2_{\mu, \alpha\ast, \rm{H}}$ by $\varepsilon^2_{\mu, \alpha\ast} +
c^2_\mu$, and $\varepsilon^2_{\mu, \delta, \rm{H}}$ by $\varepsilon^2_{\mu,
\delta, \rm{H}} + c^2_\mu$, respectively. No changes are made in the
non-diagonal
elements (except for decoupling the parallax as described above), i.e. all the
covariances remain in the LTP as they were in the SI mode, e.g.
\begin{equation}
D_{\rm{H, 12, LTP}} = D_{\rm{H, 12}} = \varepsilon_{\alpha\ast, \rm{H}}
(T_{\rm H}) \varepsilon_{\delta, \rm{H}} (T_{\rm H}) \rho_{\alpha\delta,
\rm{H}} (T_{\rm H}) \,\, .                                             %(61)
\end{equation}

%Section 4
\section{Short-term prediction (STP)}

In this section, we consider the other limiting case of the general solution
$x_{\rm p} (t)$, namely the limit for $\Delta t = t - T_2 \rightarrow 0$. This
`short-term prediction' $x_{\rm{STP}} (t)$ is valid for epochs close to $T_2$
(in the case of using HIPPARCOS: $T_2 \sim 1991.25$).

For the case $\Delta t \rightarrow 0$, we use for the correlation function
$\xi (\Delta t)$ and $\zeta (\Delta t)$ Taylor series in $\Delta t$, and keep
only terms which are linear in $\Delta t$.
From the Eqs. (54) and (55) of WPSA, we obtain
for small values of $\Delta t = t - T_2$
\begin{eqnarray}
\xi (\Delta t )  & \sim & \xi (0) \,\, ,                             \\ %(62)
\zeta (\Delta t) & \sim & \eta (0) \Delta t \,\, .                      % (63)
\end{eqnarray}
In Eq. (63), we have made use of the differential relation (28) of WPSA.
Although HIPPARCOS values are already averaged over about 3 years of
observation, the use of Eq. (63) is justified according to numerical
investigations carried out by M. Biermann (1996, unpublished, see also WPSA,
Sect. 3.4).

Inserting Eqs. (62) and (63) into the Eqs. (22) and (23), we find for the
short-term prediction
\begin{eqnarray}
\lefteqn{
\gamma_{\rm{STP}} (t) = \Big((\xi (0) + \varepsilon^2_{x, 1}) (\eta (0) +
\varepsilon^2_{\mu, 1} + \varepsilon^2_{\mu, 2})
}                                                              \nonumber\\
& & + \, \varepsilon^2_{\mu, 1} (T_2 - T_1)^2 (\eta (0) +
    \varepsilon^2_{\mu, 2})                                     \nonumber\\
& & + \, \varepsilon^2_{\mu, 1} \varepsilon^2_{\mu, 2} (T_2 - T_1)
    (t - T_2)\Big) / N \,\, ,                                 \\[0.5ex] %(64)
\lefteqn{
B_{\rm{STP}} (t) = \beta_{\rm{STP}} (t) (t - T_2)
}                                                            \nonumber\\
& & = \bigg(\varepsilon^2_{\mu, 1} \varepsilon^2_{x, 2} (T_2 - T_1)
                                                              \nonumber\\
& & \hspace{0.5 cm} + \Big(\eta (0) \big(\xi (0) + \varepsilon^2_{x, 1} +
    \varepsilon^2_{x, 2} + \varepsilon^2_{\mu, 1} (T_2 - T_1)^2 \big)
                                                              \nonumber\\
& & \hspace{1cm} + \, \varepsilon^2_{\mu, 1}
    \big( \xi (0) + \varepsilon^2_{x, 1} +
    \varepsilon^2_{x, 2} \big )\Big) (t - T_2)\bigg) / N \,\, ,         %(65)
\end{eqnarray}
where $N$ is still given by Eq. (24). Inserting $\gamma_{\rm{STP}} (t)$ and
$B_{\rm{STP}} (t)$ into Eq. (1), we obtain
\begin{equation}
x_{\rm{STP}} (t) = x_{\rm{STP}} (T_2) + \mu_{\rm{STP}} \, (t - T_2) \,\, ,%(66)
\end{equation}
which is a linear function of the epoch $t$. We do not use here a `central'
epoch $T_{\rm{STP}}$, since our Taylor series for $x_{\rm{STP}} (t)$ is
centered around
$t = T_2$. The proper motion $\mu_{\rm{STP}}$ can be derived as the sum of the
coefficients in front of $t$ in the Eqs. (1), (63), and (64):
\begin{eqnarray}
\lefteqn{
\mu_{\rm{STP}} = \mu_1    }                                 \nonumber\\
\lefteqn{
+ \,  \Big( \big( x_2 (T_2) - x_1 (T_1) - \mu_1 \, (T_2 - T_1) \big) \,
\varepsilon^2_{\mu, 1} \varepsilon^2_{\mu, 2} \, (T_2 - T_1) \Big) / N
                          }                                         \nonumber\\
\lefteqn{ + \, (\mu_2 - \mu_1) \Big(\eta (0) \big( \xi (0) +
    \varepsilon^2_{x, 1} + \varepsilon^2_{x, 2} + \varepsilon^2_{\mu, 1}
    (T_2 - T_1)^2 \big)   }                                 \nonumber\\
& & \hspace{3.0 cm} + \,\, \varepsilon^2_{\mu, 1}
\big( \xi (0) + \varepsilon^2_{x, 1}
    + \varepsilon^2_{x, 2}\big) \Big) / N \,\, .                        %(67)
\end{eqnarray}
After some algebra, $\mu_{\rm{STP}}$ can be rewritten as
\begin{equation}
\mu_{\rm{STP}} = \frac{w_{\mu, 10, \rm{STP}} \, \mu_{10}
+ w_{\mu, 2} \, \mu_2}
{w_{\mu, 10, \rm{STP}} + w_{\mu, 2}} \,\, .                           %(68)
\end{equation}
The `combined' mean proper motion $\mu_{10}$ is derived from $\mu_1$ and
$\mu_0$ by
\begin{equation}
\mu_{10} = \frac{w_{\mu, 1} \, \mu_1 + w_{\mu, 0, \rm{LTP}} \, \mu_0}
{w_{\mu, 1} + w_{\mu, 0, \rm{LTP}}}  \,\, .                           %(69)
\end{equation}
$\mu_0, \, w_{\mu, 1}$, and $w_{\mu, 2}$ are given by the corresponding
equations in Sect. 2.4, and $w_{\mu, 0, \rm{LTP}}$ by Eq. (53). The weight
$w_{\mu, 10, \rm{STP}}$ of $\mu_{10}$ in the STP solution is given by
\begin{equation}
w_{\mu, 10, \rm{STP}} = \frac{1}{(1/w_{\mu, 10, \rm{LTP}}) + c^2_\mu}
= \frac{1}{\varepsilon^2_{\mu, 10, \rm{LTP}} + c^2_\mu} \,\, ,        %(70)
\end{equation}
with
\begin{equation}
w_{\mu, 10, \rm{LTP}} = w_{\mu, 1} + w_{\mu, 0, \rm{LTP}} = \frac{1}
{\varepsilon^2_{\mu, 10, \rm{LTP}}} \,\, .                            %(71)
\end{equation}
The meaning of Eqs. (68)-(71) is the following: $\mu_{\rm{STP}}$ is the
weighted
mean of $\mu_2$ and the combined mean proper motion $\mu_{10}$, where the
cosmic error $c_\mu$ in $\mu_{10}$ has to be taken into account in $w_{\mu,
10, \rm{STP}}$. The combined mean proper motion $\mu_{10}$ is itself a weighted
mean of $\mu_1$ and $\mu_0$, where for $\mu_0$ the cosmic error $c_x$ in
$x_2 (T_2)$
has to be included into $w_{\mu, 0, \rm{LTP}}$.

The exact expression for $x_{\rm{STP}} (T_2) $ is given by
\begin{eqnarray}
\lefteqn{
x_{\rm STP} (T_2) = x_2 (T_2)
}                                                           \nonumber\\[0.5ex]
\lefteqn{
- \, \varepsilon^2_{x, 2} \Big( \big(\eta (0)
       + \varepsilon^2_{\mu, 1} + \varepsilon^2_{\mu, 2}) (x_2 (T_2)
       - x_1 (T_1) - \mu_1 (T_2 - T_1) )                }   \nonumber\\[0.5ex]
\lefteqn{
\hspace{+ 2.50 cm}
- \, \varepsilon^2_{\mu, 1}
      (T_2 - T_1) (\mu_2 (T_2) - \mu_1) \Big) / N \,\, .   }          %(72)
\end{eqnarray}
For all practical purposes we can, however, neglect the last term in Eq.
(72), which is proportional to $\varepsilon^2_{x, 2}$. The measuring accuracy
$\varepsilon_{x, 2}$ for the HIPPARCOS positions $x_2 (T_2)$ at $T_2 \sim
1991.25$ is so high, relative to the other measuring errors and cosmic errors,
that this term is usually of the order of 0.01 mas only. We use
instead of Eq. (72) the very good approximation
\begin{equation}
x_{\rm{STP}} (T_2) = x_2 (T_2)  \,\, .                                 %(73)
\end{equation}
The mean errors of $x_{\rm{STP}} (T_2)$ and of $\mu_{\rm{STP}}$ are given by
\begin{equation}
\varepsilon^2_{x, \rm{STP}} (T_2)   =
\varepsilon^2_{x, 2} (T_2) \,\, ,                                      %(74)
\end{equation}
\begin{equation}
\varepsilon^2_{\mu, \rm{STP}}   =   \frac{1}
{w_{\mu, 10, \rm{STP}} + w_{\mu, 2}}   \,\, .                          %(75)
\end{equation}
The full uncertainty $\varepsilon_{x, \rm{STP}} (t)$ of $x_{\rm{STP}} (t)$ is
given by
\begin{eqnarray}
\lefteqn{
\varepsilon^2_{x, \rm{STP}} (t) = \varepsilon^2_{x, \rm{STP}} (T_2) +
\varepsilon^2_{\mu, \rm{STP}} \, (t - T_2)^2
}                                                              \nonumber\\
& & \hspace{1.4 cm} + \, \frac{1}{4} \, \xi^{(IV)}_0 \, (t - T_2)^4 \,\,. %(76)
\end{eqnarray}
In Eq. (76), we have neglected the very small correlation between $x_{\rm{STP}}
(T_2) = x_2 (T_2)$ and $\mu_{\rm{STP}}$ which is caused by the use of $x_2
(T_2)$
in deriving $\mu_0$ which in turn enters into $\mu_{\rm{STP}}$. The last term
in Eq. (76)
is the statistical uncertainty of a prediction based on instantaneous data
at $T_2$ for small epoch differences $\Delta t$ (see Eq. (57) of WPSA).
$\xi^{(IV)}_0$ is the fourth derivative of the correlation function
$\xi (\Delta t)$ with respect to $\Delta t$ at $\Delta t = 0$.

In the numerical approach for deriving the short-term prediction, we modify the
procedure for the single-star mode presented in Paper I more strongly than for
the LTP solution. As `observations' {\bf b} we use first {\bf b}$_{\rm H}
(T_{\rm H})$, as
in Paper I, the parallax $p_{\rm H}$ inclusive. The corresponding part {\bf
D}$_{\rm H}$ of
the variance-covariance matrix {\bf D} remains also unchanged. The second part
of {\bf b}, which we now call {\bf b}$_{\rm m}$ is given by the
$\alpha_\ast$ and
$\delta$ components of the combined mean proper motion $\mu_{10}$. The quantity
$\mu_{10}$ is obtained in a preparatory step from Eq. (69), and its mean error
from Eq. (70) as $(w_{\mu, 10, \rm{STP}})^{-1/2}$. Each component of
$\mu_{10}$,
i.e. of {\bf b}$_{\rm m}$, is considered not to be correlated with any
other component of {\bf b}. The numerical approach for STP produces values for
$x_{\rm{STP}}$ and $\mu_{\rm{STP}}$ (in $\alpha_\ast$ and $\delta$), a new
parallax
$p_{\rm{STP}}$, and the corresponding variance-covariance matrix. In presenting
the
results in printed form, we use again central epochs $T_{\rm{STP}}$ (different
for
$\alpha_\ast$ and $\delta$), at which $x_{\rm{STP}}$ and
$\mu_{\rm{STP}}$
are uncorrelated. However, $T_{\rm{STP}}$ differs usually only very slightly
from the individual central epochs of the basic HIPPARCOS data.

%Section 5
\section{Transition from the short-term prediction to the long-term prediction}

%Section 5.1
\subsection{Transition in position}

As discussed in Sect. 2.4 and illustrated in Fig. 2, the general solution
$x_{\rm p} (t)$ is a smooth transition from the short-term prediction
$x_{\rm{STP}} (t)$
for epochs $t$ close to $T_2$ to the long-term prediction $x_{\rm{LTP}} (t)$
for
$t \rightarrow \pm \, \infty$. We are now asking for the `transition function'
$\beta_{\rm{trans}} (t)$ which describes this transition in $x$:
\begin{equation}
x_{\rm p} (t) = (1 - \beta_{\rm{trans}} (t)) \, x_{\rm{LTP}} (t) +
\beta_{\rm{trans}} (t) \, x_{\rm{STP}} (t) \,\, .                        %(77)
\end{equation}
Formally we can always solve Eq. (77) for $\beta_{\rm{trans}} (t)$, using our
former results for $x_{\rm p} (t), \, x_{\rm{LTP}} (t),$ and $x_{\rm{STP}}
(t)$. However, the
resulting transition function $\beta_{\rm{trans}} (t)$ is then very complicated
and depends unfortunately not only on the correlation functions, but also
explicitely on the measured values of $x_1, x_2, \mu_1, \mu_2$, and on their
measuring errors. Such a result is not very suitable for a practical
application.

There does exist, however, an approximate treatment for the transition function
$\beta_{\rm{trans}} (t)$ which gives a very simple and easily applicable
result, and which nevertheless describes the transition quantitatively rather
accurately. The basic idea is the observation that in real applications the
mean position $x_1 (T_1)$ enters into the final result $x_{\rm p} (t)$ mainly
through
the proper motion $\mu_0$. This is caused by the small error of the HIPPARCOS
position $x_2 (T_2)$ with respect to the error of the mean (FK5 or GC) position
$x_1 (T_1)$. Only in cases of a large cosmic error $c_x (p)$ in $x_2 (T_2)$,
our approximation becomes less accurate. We therefore consider the transition
function $\beta_{\rm{trans}} (t)$ for the limiting case in which
$\varepsilon_{x, 1}$ tends towards infinity while $\varepsilon_{\mu, 0}$
remains finite (equal to its actual value). The latter can be enforced by
setting $T_2 - T_1 = \varepsilon_{x, 1} / \varepsilon_{\mu, 0}$. This means
that we let go $T_1 \rightarrow - \, \infty$ and $\varepsilon_{x, 1}
\rightarrow + \, \infty$ in such a way that $\varepsilon_{\mu, 0}$ remains
constant.

%%%%%%%%%%%%%%%%%%%%%%%%%%%%%%%%% Begin of Figure 2 %%%%%%%%%%%%%%%%%%%%%%
% Begin of        %
%                 %
% F i g u r e   2 %
%                 %
%%%%%%%%%%%%%%%%%%%
\begin{figure}[t]
\resizebox{\hsize}{!}{\includegraphics{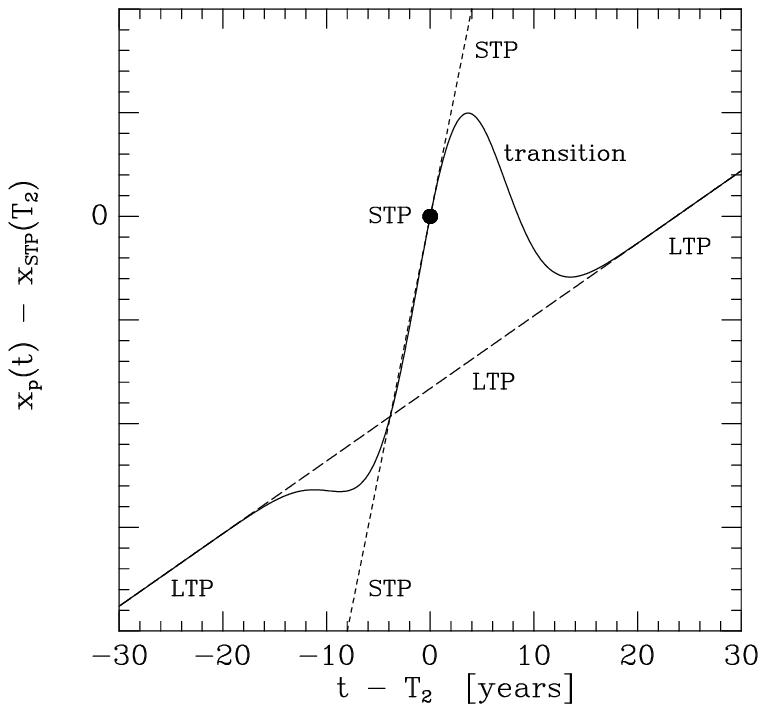}}
\caption[]
{Schematic illustration of the non-linear transition
of the 'best' prediction $x_{\rm p}(t)$ (solid curve)
from the
linear short-term prediction $x_{\rm {STP}}(t)$
(STP mode, valid for epochs $t \sim T_2$; short-dashed line)
to the linear long-term prediction $x_{\rm {LTP}}(t)$
(LTP mode, valid for epochs $|t - T_2|  \rightarrow \infty $;
long-dashed line).}
\end{figure}
%%%%%%%%%%%%%%%%%%%%%%%%%%%%%%%% End of Figure 2 %%%%%%%%%%%%%%%%%%%%%%%%%%%%%

%%%%%%%%%%%%%%%%%%%
% Begin of        %
%                 %
% T a b l e    1  %
%                 %
%%%%%%%%%%%%%%%%%%%
\begin{table}[t]
\caption{
Transition functions $\beta_{\rm{trans}} (t)$ and $\nu_{\rm{trans}} (t)$
for a simple example
}
\begin{tabular}{ccc}
\hline\\[-1.5ex]
$|\Delta t| = |t - T_2|$ & $\beta_{\rm{trans}}(\Delta t)$ & $\nu_{\rm{trans}}
(\Delta t)$\\
$\left[\rm{years}\right]$ & \\[0.5ex]\hline\\[-1.4ex]
0  & 1.000 & \hphantom{--} 1.000\\
1  & 0.978 & \hphantom{--} 0.934\\
2  & 0.915 & \hphantom{--} 0.752\\
3  & 0.819 & \hphantom{--} 0.491\\
4  & 0.700 & \hphantom{--} 0.202\\
5  & 0.573 &           --  0.065\\
6  & 0.449 &           --  0.270\\
7  & 0.336 &           --  0.397\\
8  & 0.241 &           --  0.445\\
9  & 0.165 &           --  0.430\\
10\hphantom{1} & 0.108 & -- 0.373\\
11\hphantom{1} & 0.068 & -- 0.297\\
12\hphantom{1} & 0.041 & -- 0.219\\
13\hphantom{1} & 0.023 & -- 0.152\\
14\hphantom{1} & 0.013 & -- 0.099\\
15\hphantom{1} & 0.007 & -- 0.060\\[0.5ex]
20\hphantom{2} & \hphantom{1}0.0001 & -- 0.002\\[0.5ex]\hline
\vspace*{-0.55cm} %%%%%%%%%%%%%%%%%%%%%%%%%%%%%%%%%%%%%%%%%%%%%%%%%%%%%%%%%%%
\end{tabular}
\end{table}
%%%%%%%%%%%%%%%%%%%%%%%%%%%%%%%% End of Table 1 %%%%%%%%%%%%%%%%%%%%%%%%%%%%%

If we use this special case as an approximation, we derive after some lengthy
algebra the following rather simple expression for the transition function:
\begin{equation}
\beta_{\rm{trans}} (t) = \frac{\zeta (t - T_2)}
{\eta (0) \, (t - T_2)} \,\, .                                            %(78)
\end{equation}
A similar function has already been derived as Eq. (70) in Sect. 4.2.4 of WPSA.
The function $\beta_{\rm{trans}}$ has the welcome property that it does neither
depend on the measured values of $x$ and $\mu$ nor on the mean errors of a
given star. These values are fully absorbed in the individual solutions
$x_{\rm{STP}} (t)$ and $x_{\rm{LTP}} (t)$. Hence $\beta_{\rm{trans}} (t)$ is
the same
function for all the objects. We should remark here that the apparently more
complicated form of the transition function according to Eq. (75) of WPSA is
caused by the fact that we have used in WPSA a slightly different definition of
the long and short-term prediction, namely $x_2 (t)$ instead of $x_{\rm{STP}}
(t)$ and $x_1 (t)$ instead of $x_{\rm{LTP}} (t)$.

The function $\beta_{\rm{trans}} (t)$ has the desired properties in the limits
$\Delta t \rightarrow 0$ or $\infty$. For $t = T_2$, we have
$\beta_{\rm{trans}} (T_2) = 1$,
because of $\zeta(\Delta t) \sim \eta(0) \Delta t$
for small values of $\Delta t$.
For $t
\rightarrow \infty, \, \zeta (t - T_2)$ vanishes  and hence
$\beta_{\rm{trans}} (\infty) = 0$. An example for the full run of
$\beta_{\rm{trans}} (t - T_2)$, from 1 to 0, is illustrated by the full curve
shown in Fig. 8 of WPSA. Table 1 gives $\beta_{\rm{trans}} (t -
T_2)$ for a few values of $|t - T_2|$, using the example given in
Sect. 3.6 of WPSA for the correlation function $\zeta
(\Delta t)$. At $|t - T_2| \sim$ 5-6 years,
the general solution $x_{\rm p} (t)$
is
about half-way between $x_{\rm{STP}} (t)$ and $x_{\rm{LTP}} (t)$. From both the
Fig. 8 of WPSA and Table 1,
we get an indication for the range of applicability of the
short and long-term prediction. The short-term prediction $x_{\rm{STP}} (t)$
has
a rather limited range of applicability, namely a few years around $T_2$ only.
The long-term prediction $x_{\rm{LTP}} (t)$ is a good approximation for the
general
solution $x_{\rm p} (t)$ for epoch differences $|t - T_2|$ which are larger
than
about 10 years. Hence the transition from STP to LTP is rather rapid, at least
in our example.

We should remark here that (by chance) the transition function $\beta_{\rm
trans} (t)$ given by Eq. (78) is even strictly
valid (exact) for the example used for the correlation functions in Sect. 3.6
of WPSA and adopted in Table 1. For more general runs of the correlation
functions, $\beta_{\rm trans}$ can become quite large
at epochs around the `crossing time' at which $x_{\rm LTP} = x_{\rm STP}$, if
$x_{\rm LTP}$ and $x_{\rm STP}$ are significantly different from $x_{\rm p}$
at that time.

%Section 5.2
\subsection{Transition in proper motion}

The most accurate prediction $\mu_{\rm p} (t)$ for the instantaneous proper
motion $\mu (t)$ at an arbitrary epoch $t$
is formally given by
\begin{equation}
\mu_{\rm p} (t) = \dot{x}_{\rm p} (t) = (1 - \nu_{\rm{trans}}
(t))\,\mu_{\rm{LTP}} + \nu_{\rm{trans}} (t)\,\mu_{\rm{STP}} \,\, .    %(79)
\end{equation}
The transition function $\nu_{\rm{trans}} (t)$ for the proper motion $\mu$ is,
similar to $\beta_{\rm{trans}}$, in general a rather complicated function,
which depends on the measured values of $\mu_1, \mu_2,$ and $\mu_0$. If we
adopt the same approximation as used in Sect. 5.1 for obtaining Eq. (78) for
$\beta_{\rm{trans}}$, namely $\varepsilon_{x, 1} \rightarrow \infty$ while
keeping $\varepsilon_{\mu, 0}$ constant, we derive for $\nu_{\rm{trans}} (t)$
also a very simple expression:
\begin{equation}
\nu_{\rm{trans}} (t) = \beta_{\rm{trans}} (t) + \dot{\beta}_{\rm{trans}} (t)
(t - T_{\rm{STP}}) = \frac{\eta (t - T_{\rm{STP}})}{\eta (0)} .        %(80)
\end{equation}
This transition function $\nu_{\rm{trans}} (t)$ for $\mu (t)$ has the same
proper limiting values as $\beta_{\rm{trans}} (t)$: For $t \rightarrow \infty$,
we have $\nu_{\rm{trans}} (\infty) = 0$ and hence $\mu_{\rm p} (\infty) =
\mu_{\rm{LTP}}$, and for $t = T_{\rm{STP}}$ we obtain $\nu_{\rm{trans}}
(T_{\rm{STP}}) = 1$ and $\mu_{\rm p} (T_{\rm{STP}}) = \mu_{\rm{STP}}$ .
In Table 1 we list $\nu_{\rm{trans}} (\Delta t)$ for a few values of
$|\Delta t| = |t - T_{\rm{STP}}|$, using Eq. (80) and the simple example
for $\eta (\Delta t)$ given in Sect. 3.6 of WPSA.

The transition functions $\beta_{\rm{trans}} (t)$ and $\nu_{\rm{trans}} (t)$
according to Eqs. (78) and (80) have another nice property: If we use the
example given in Sect. 3.6 of WPSA, then the function $\beta_{\rm{trans}}$
depends only on $t - T_2$, but not on the individual cosmic error $c_\mu (p) =
(\eta (0, p))^{1/2}$, since $\eta (0)$ occurs also as a factor in $\zeta (t -
T_2)$ and cancels out in Eq. (78). Similarly, the transition function
$\nu_{\rm{trans}} (t)$ is a `scaled' function and depends on $t - T_2$ only,
but not on $\eta (0, p)$.

%Section 6
\section{Cosmic errors}

For deriving the solutions in the LTP and STP mode, we need to know the cosmic
errors $c_\mu = (\eta (0))^{1/2}$ and $c_x = (\xi (0))^{1/2}$. The cosmic
errors depend strongly on the distance $r$ of the star from the Sun, or
equivalently on the stellar parallax $p$. For 1202 `apparently single stars'
from the basic FK5, we have obtained the functions $c_\mu (p)$ and $c_x (p)$
empirically by using groups of stars in various distance intervals. The
data can be represented by the following fit functions:
\begin{eqnarray}
c_\mu (p) & = \, (\eta (0, p))^{1/2} & = \, \left( \frac{C_1\,p}{(C^2_2 +
p^2)^{1/2}}\right)^{1/2} \,\, , \\[1ex]                                 %(81)
c_x (p) & = \, (\xi (0, p))^{1/2} & = \,\, C_3\,c_\mu (p)\,\, ,       %(82)
\end{eqnarray}
with
\begin{eqnarray}
C_1 & = & 9.30 \,\,\rm{(mas/year)}^2 \,\, , \\                          %(83)
C_2 & = & 22.14 \,\,\rm{mas}           \,\, , \\                        %(84)
C_3 & = & 5.93 \,\,\rm{years}        \,\, .                             %(85)
\end{eqnarray}
These versions of $c_\mu (p)$ and $c_x (p)$ have been used for the FK6 (Wielen
et al. 1999d, 2000a).
A table for $c_\mu$ and $c_x$ as a function of $p$ or $r$,
based on Eqs. (81)-(85), is given on page 12 of Wielen et al. (1999d).

In the future we hope to determine also the dependence of the cosmic errors on
the absolute magnitude (or mass) of the stars for a given parallax. A
comparison of the results
for the cosmic errors based on the FK5 stars with those derived from the (on
average fainter) GC stars seems to indicate a rather weak dependence on the
brightness of the stars (Wielen et al. 1998).

%%%%%%%%%%%%%%%%%%%
% Begin of        %
%                 %
% T a b l e  2    %
%                 %
%%%%%%%%%%%%%%%%%%%

\begin{table*}[t]
\caption
{
Results of the combination method for the star HIP 9884 = FK 74 = GC 2538 =
$\alpha$ Ari
for the various modes
}
\begin{tabular}{lrrrrrrrrrrrr}
\hline\\[-1.0ex]
& \multicolumn{6}{c}{FK5 + HIPPARCOS} & \multicolumn{6}{c}{GC + HIPPARCOS}\\
Quantity & $\alpha_\ast$ & \multicolumn{2}{c}{mean error} &
$\delta$ & \multicolumn{2}{c}{mean error} & $\alpha_\ast$ &
\multicolumn{2}{c}{mean error} & $\delta$ & \multicolumn{2}{c}{mean error}
\\[0.5ex]\hline\\[-1.8ex]
Input data:
\hspace{3.00 cm}
\\[1ex]
$\Delta x_{\rm F} (T_{\rm F})$ or $\Delta x_{\rm{GC}} (T_{\rm{GC}})$
& -- 7.83 & $\pm$ 12.52 && +
96.95 & $\pm$ 15.30 && -- 2.64 & $\pm$ 49.92 && + 236.68 & $\pm$ 32.88 &\\
$\Delta\mu_{\rm F}$ or $\Delta\mu_{\rm{GC}}$
& + 0.49 & $\pm$\hphantom{1} 0.40 &&
--\hphantom{1} 1.20 & $\pm$\hphantom{1} 0.38 && + 1.69 &
$\pm$\hphantom{3} 1.24 && --\hphantom{23} 3.98 & $\pm$\hphantom{2} 1.86 &\\
$T_{\rm F}$ or $T_{\rm{GC}}$
& 1947.84 & & & 1929.73 & & & 1892.60 && & 1890.30
&&\\[1.5ex]
$\Delta x_{\rm H} (T_{\rm {H, ind}})$
& 0.00 & $\pm$ 0.77 && 0.00 & $\pm$ 0.54 && 0.00 &
$\pm$ 0.77 && 0.00 & $\pm$ 0.54 &\\
$\Delta\mu_{\rm H}$
& 0.00 & $\pm$ 1.01 && 0.00 & $\pm$ 0.77 && 0.00 & $\pm$ 1.01 &&
0.00 & $\pm$ 0.77 &\\
$T_{\rm{H, ind}}$
& 1991.26 & && 1991.51 && & 1991.26 && & 1991.51 &&\\[1.5ex]
$p_{\rm H}$ & 49.48
& $\pm$ 0.99 &&& & & 49.48 & $\pm$ 0.99 &&& &\\[1.5ex]
$c_x$
& 17.28 & & & 17.28 & & & 17.28 & & & 17.28 & &\\
$c_\mu$
&  2.91 & & &  2.91 & & &  2.91 & & &  2.91 & &\\[2ex]
\hline\\[-1ex]
$\Delta\mu_0$ & + 0.18 & $\pm$ 0.29 && -- 1.57 & $\pm$ 0.25 && + 0.03 & $\pm$
0.51 && -- 2.34 & $\pm$ 0.32 &\\[1.5ex]
\hline\\[-1ex]
\multicolumn{5}{l}{
Results for the single-star mode {\bf SI}:
}
\\[1ex]
$\Delta x_{\rm {SI}} (T_{\rm {SI}})$
& -- 0.18 & $\pm$ 0.76 && + 0.14 & $\pm$ 0.54 && -- 0.23 &
$\pm$ 0.77 && + 0.08 & $\pm$ 0.54 &\\
$\Delta\mu_{\rm {SI}}$
& + 0.23 & $\pm$ 0.23 && -- 1.34 & $\pm$ 0.20 && + 0.04 & $\pm$
0.42 && -- 2.00 & $\pm$ 0.29 &\\
$T_{\rm {SI}}$
& 1991.12 && & 1991.47 && & 1991.26 && & 1991.51 &&\\[1.5ex]
$\Delta p_{\rm {SI}}$
& -- 0.57 & $\pm$ 0.95 & & &&& -- 0.98 & $\pm$ 0.95 &&&
&\\[0.5ex]
\hline\\[-1ex]
\multicolumn{5}{l}{
Results for the long-term prediction mode {\bf LTP}:
}
\\[1ex]
$\Delta x_{\rm {LTP}} (T_{\rm {LTP}})$
& -- 5.14 & $\pm$ 10.14 && + 54.37 & $\pm$ 11.46
&& -- 0.30 & $\pm$ 16.34 && + 51.26 & $\pm$ 15.30 &\\
$\Delta\mu_{\rm {LTP}}$
& + 0.36 & $\pm$ 0.31 && -- 1.38 & $\pm$ 0.27
&& + 0.28 & $\pm$ 0.49 && -- 2.37 & $\pm$ 0.36 &\\
$T_{\rm {LTP}}$
& 1962.77 && & 1956.76 && & 1980.68 && & 1969.40 &&\\[2ex]
\hline\\[-1ex]
\multicolumn{5}{l}{
Results for the short-term prediction mode {\bf STP}:
}
\\[1ex]
$\Delta x_{\rm {STP}} (T_{\rm {STP}})$
& -- 0.01 & $\pm$ 0.77 &&  0.00 & $\pm$ 0.54
&& -- 0.02 & $\pm$ 0.77 &&  0.00 & $\pm$ 0.54 &\\
$\Delta\mu_{\rm {STP}}$
&  0.00 & $\pm$ 0.95 && -- 0.08 & $\pm$ 0.74
&& -- 0.03 & $\pm$ 0.95 && -- 0.15 & $\pm$ 0.74 &\\
$T_{\rm {STP}}$
& 1991.26 && & 1991.52 && & 1991.26 && & 1991.52 &&\\[1.5ex]
$\Delta p_{\rm {STP}}$
& -- 0.03 & $\pm$ 0.99 & & &&& -- 0.06 & $\pm$ 0.99 &&&
&\\[0.5ex]\hline\\[-1ex]
\end{tabular}
Units: mas, mas/year, or years
\end{table*}

%%%%%%%%%%%%%%%%%%%%%%%%%%%%%% End of Table 2 %%%%%%%%%%%%%%%%%%%%%%%%%%%%%

%Section 7
\section{An example: $\alpha$ Ari}

In order to illustrate our combination method for all the three modes (SI, LTP,
STP), we give in Table 2 the results for one individual star. As in Paper I
(Sect. 5), we give the positions $x (t)$ and the proper motions $\mu (t)$
always relative to the HIPPARCOS solution as $\Delta x (t) = x (t) - x_{\rm H}
(t)$ and $\Delta\mu (t) = \mu (t) - \mu_{\rm H} (t)$,
in order to save printing space and
to make the comparison of the results easier. The ground-based data are always
reduced to the HIPPARCOS system. The results of the single-star mode presented
here in Paper II differ somewhat from those of Paper I, because we now adopt
slightly improved systematic differences FK5-HIP and GC-HIP. For a valid
comparison of the results of the three different modes it is necessary to use
exactly the same basic input data.

Table 2 shows that the short-term prediction is usually quite close to the
HIPPARCOS solution. On the other hand, the long-term prediction differs most
strongly from the HIPPARCOS solution, since the HIPPARCOS values are entering
with a lower weight into the LTP mode than in the SI mode, because of the
cosmic errors in the HIPPARCOS data. The central epoch $T_{\rm{LTP}}$ is the
only one which is usually significantly earlier than $T_{\rm H} \sim 1991.25$,
and the mean error of the central position $x_{\rm{LTP}} (T_{\rm{LTP}})$ is
typically only slightly smaller than the cosmic error $c_x (p)$.

%Section 8
\section{Error budget}

In Table 3 we present the error budget of proper motions in the three different
modes (SI, LTP, STP) for two samples of basic FK5 stars. The mean errors
$\varepsilon_\mu$ given in Table 3 refer to one `mean' coordinate component. It
is obtained as an rms average over $\varepsilon_{\mu, \alpha\ast}$ and
$\varepsilon_{\mu, \delta}$, and over all the stars in the
corresponding sample. The error budget for the 1535 basic FK5 stars is slightly
fictious, since this sample contains double stars for which the FK6 provides in
reality `special' solutions instead of the `direct' combination solutions
discussed in this paper. Nevertheless, the results for this sample provide a
valid indication for the overall accuracy of our combination method in the
direct modes SI, LTP, and STP. The sample of 1202 basic FK5 stars contains
`apparently single objects' only. Most of these stars (878 objects) have direct
solutions in the FK6. The error budget for these 878 basic FK5 stars in Part I
of the FK6 is given in Wielen et al. (1999d). The error budgets for 3272
additional fundamental stars with direct solutions in the three modes are
presented in Part III of the FK6 (Wielen et al. 2000a).

Table 4 gives the error budget for the combination of the GC (Boss et al. 1937)
with HIPPARCOS. The sample of GC stars are the `full sample' of 29\,717 GC
stars observed by HIPPARCOS, and the `subsample' of 11\,737 GC stars with
linear HIPPARCOS standard solutions.
From Table 4 it is obvious that the original proper motions
$\mu_{\rm {GC}}$
of the GC do not contribute significantly to the GC+HIP on average.
However, for some brighter and well-observed stars, the accuracy of
$\mu_{\rm {GC}}$
is much better than the rms value of
$\varepsilon_{\mu, {\rm GC}}$ seen in Table 4.
Furthermore, the proper motion
$\mu_{0{\rm {(GC)H}}}$,
derived from the central positions of the GC and the HIPPARCOS Catalogue,
has usually a quite reasonable accuracy for most of the GC stars.

The typical gain in accuracy in the proper motions derived in the long-term
prediction mode, relative to the original HIPPARCOS proper motions, is a
factor of 4.6 for the basic fundamental stars in the FK6\,=\,FK5+HIP, and a
factor of 1.8 for the 11\,773 GC stars in the GC+HIP. This improvement in the
LTP mode with respect to HIPPARCOS is a consequence of the cosmic errors in the
instantaneously  measured HIPPARCOS proper motions. In contrast to the LTP
mode, the short-term predictions (STP mode) do not differ so much from the
HIPPARCOS solutions. However, in most cases we are more interested in the
long-term averaged proper motion (LTP) or in the single-mode result (SI), where
the gain in accuracy is quite significant.

%Section 9
\section{Problems and applications}

The long-term predictions and the short-term predictions derived in the former
sections are {\it statistically} the best astrometric solution for a sample of
`apparently single-stars'. However, an unsatisfactory property of these
statistically valid solutions is the fact that we are not able to make proper
use of the available information on the {\it individual} behaviour of the
stars. For example, we use the overall cosmic errors $c_\mu (p)$ and $c_x (p)$,
no matter whether the object is a $\Delta\mu$ binary or a single-star candidate
(Wielen et al. 1999a).
In principle, for each individual star, one would have to use
`conditioned correlation functions' which are based on the individually
observed differences between the instantaneous measurements and the mean data
(e.g. on $\mu_{\rm{FK5}} - \mu_{\rm{HIP}}$). Unfortunately, the conditioned
correlation functions, i.e. `conditioned cosmic errors' in particular, are not
available at present.

We should also point out that presently the cosmic errors $c_x$ in position are
much more uncertain than the cosmic errors $c_\mu$ in proper motion. While the
typical values of $c_\mu$ are larger than the measuring errors
$\varepsilon_{\mu, \rm{FK5}}$, $\varepsilon_{\mu, 0}$, and
$\varepsilon_{\mu, \rm{HIP}}$, the typical values of $c_x$ are nearly lost in
the measuring errors of the ground-based data.

Furthermore, our knowledge about the actual form of the correlation functions
$\xi (\Delta t)$, $\eta (\Delta t)$, $\zeta (\Delta t)$ for
$\Delta t \!\!= \!\!\!\!\!\!\!\!|\,\,\,\,0$
is presently still quite rudimentary. The simple example for the correlation
functions presented in Sect. 3.6 of WPSA is mainly given for illustrating the
general behaviour of these functions. The application of this example to real
data should be done very cautiously. Hence the real transition from the STP
solution to the LTP solution (discussed in Sect. 5) is quantitatively not
well-determined.

Which of the three solutions offered (SI, LTP, STP) should be used in real
applications\,? For single-star candidates, the single-star mode should be
adopted, although some of the single-star candidates may nevertheless be
binaries. If the user is handling a sample of `apparently single stars' (which
usually contains single-star candidates, $\Delta\mu$ binaries, and
intermediate cases), then
the LTP or STP solutions are recommended, depending on the
corresponding epoch difference $\Delta t = t - T_{\rm{H}}$ (with
$T_{\rm{H}} \sim 1991$).

The problems discussed above are especially severe for objects detected as
$\Delta\mu$ binaries (Wielen et al. 1999a). For $\Delta\mu$ binaries the
difference between the instantaneous proper motion ($\mu_{\rm{HIP}}$) and the
mean one (e.g., $\mu_{\rm{FK5}}$ or $\mu_0$) is sometimes much larger than the
cosmic error $c_\mu$ expected on average. In such a case the weight $w_{\mu,
2, \rm{LTP}}$ (Eq. (53)) of the HIPPARCOS proper motion ($\mu_{\rm{HIP}} =
\mu_2$) is higher than appropriate for this individual object. The derived mean
proper motion $\mu_{\rm {LTP}}$ is then biased towards the HIPPARCOS proper
motion, and the derived mean error $\varepsilon_{\mu, \rm{LTP}}$ is too small.
In the case of extreme $\Delta\mu$ binaries, it is better to adopt a properly
weighted mean of $\mu_1$ (e.g. $\mu_{\rm{FK5}}$) and $\mu_0$ as a prediction
for the mean proper motion $\mu_{\rm m}$. The main problem in such a procedure
is the unknown individual value of the cosmic error in the HIPPARCOS position,
which is higher than the value of $c_x$ expected on average. This value enters
into the weight of $\mu_0$, and hence into the predicted value of
$\mu_{\rm m}$. We shall discuss this problem in more detail in a subsequent
paper. In any case, the LTP solutions (and, of course, the SI solutions) for
$\Delta\mu$ binaries are inherently the least accurate ones among the class of
direct solutions, because of the disturbing double-star nature of these
objects.

%Section 10
\section{Summary and outlook}

In this Paper II, we have derived and discussed an appropriate method to
combine a ground-based astrometric catalogue (such as the FK5 or GC) with the
HIPPARCOS Catalogue, taking cosmic errors (due to undetected binaries) in the
quasi-instantaneously measured
data into account. The method leads to long-term
predictions (LTP mode) and to short-term predictions (STP mode), which are the
limiting cases of the general solution. The general solution is a smooth
transition from the STP to the LTP mode. The case of single stars with no
cosmic errors was already treated in Paper I (SI mode). In a subsequent paper,
we shall present `special solutions' for known double stars.

%\begin{acknowledgements}
%T e x t
%\end{acknowledgements}

\newpage
{\hphantom{x}}
\newpage

%%%%%%%%%%%%%%%%%%%
% Begin of        %
%                 %
% T a b l e    3  %
%                 %
%%%%%%%%%%%%%%%%%%%
\begin{table}[t]
\caption{
Error budget for FK6 proper motions
}
\begin{tabular}{lcccc}
\hline\\[-1.0ex]
\multicolumn{5}{c}{Typical mean errors of proper motions}\\
\multicolumn{5}{c}{(in one component, averaged over $\mu_{\alpha\ast}$ and
$\mu_\delta$; units: mas/year)}\\[1.5ex]\hline\\[-2.0ex]
Sample of stars:& \multicolumn{2}{c}{1535 FK} & \multicolumn{2}
{c}{1202 FK}\\ [0.7ex]
\hline\\[-2.0ex]
& rms aver. & median & rms aver. & median\\[0.5ex]\hline\\[-1.0ex]
\multicolumn{5}{c}{\bf{SI}: \hspace*{0.3 cm} single-star mode}\\[0.0ex]
HIPPARCOS&       &      &      &     \\
\hspace*{0.5cm}random& 0.82  & 0.63 & 0.68 & 0.61\\[0.0ex]
FK5\\
\hspace*{0.5cm}random& 0.76 & 0.64 & 0.77 & 0.67\\
\hspace*{0.5cm}system& 0.28 & 0.25 & 0.28 & 0.25\\
\hspace*{0.5cm}total & 0.81 & 0.70 & 0.83 & 0.72\\[0.0ex]
$\mu_0$\\
\hspace*{0.5cm}random& 0.53 & 0.43 & 0.54 & 0.45\\
\hspace*{0.5cm}system& 0.24 & 0.23 & 0.25 & 0.23\\
\hspace*{0.5cm}total & 0.58 & 0.49 & 0.59 & 0.51\\[1.0ex]
FK6 = FK5+HIP        & 0.35 & 0.33 & 0.35 & 0.34\\[1.0ex]
ratio of HIPPARCOS& 2.3\hphantom{0} &
                    1.9\hphantom{0} & 1.9\hphantom{0} &
                    1.8\hphantom{0}\\
\hspace*{0.7cm}to FK6 errors\\[0.5ex]\hline\\[-1.0ex]
\multicolumn{5}{c}{\bf{LTP}: \hspace*{0.3 cm} long-term prediction}\\[0.0ex]
HIPPARCOS&       &      &      &     \\
\hspace*{0.5cm}random& 0.82  & 0.63 & 0.68 & 0.61\\
\hspace*{0.5cm}cosmic (in $\mu$) & 2.15  & 2.05 & 2.13 & 2.04\\
\hspace*{0.5cm}total&  2.30  & 2.14 & 2.24 & 2.13\\[0.0ex]
FK5\\
\hspace*{0.5cm}random& 0.76 & 0.64 & 0.77 & 0.67\\
\hspace*{0.5cm}system& 0.28 & 0.25 & 0.28 & 0.25\\
\hspace*{0.5cm}total & 0.81 & 0.70 & 0.83 & 0.72\\[0.0ex]
$\mu_0$\\
\hspace*{0.5cm}random& 0.53 & 0.43 & 0.54 & 0.45\\
\hspace*{0.5cm}system& 0.24 & 0.23 & 0.25 & 0.23\\
\hspace*{0.5cm}cosmic (due to $x$)   & 0.32 & 0.30 & 0.32 & 0.30\\
\hspace*{0.5cm}total & 0.66 & 0.57 & 0.68 & 0.59\\[1.0ex]
FK6 = FK5+HIP        & 0.49 & 0.44 & 0.49 & 0.45\\[1.0ex]
ratio of HIPPARCOS& 4.7\hphantom{0} &
                    4.9\hphantom{0} & 4.6\hphantom{0} &
                    4.7\hphantom{0}\\
\hspace*{0.7cm}to FK6 errors\\[0.5ex]\hline\\[-1.0ex]
\multicolumn{5}{c}{\bf{STP}: \hspace*{0.3 cm} short-term prediction}\\[0.0ex]
HIPPARCOS&       &      &      &     \\
\hspace*{0.5cm}random& 0.82  & 0.63 & 0.68 & 0.61\\[0.0ex]
FK5\\
\hspace*{0.5cm}random& 0.76 & 0.64 & 0.77 & 0.67\\
\hspace*{0.5cm}system& 0.28 & 0.25 & 0.28 & 0.25\\
\hspace*{0.5cm}cosmic (in $\mu$) & 2.15 & 2.05 & 2.13 & 2.04\\
\hspace*{0.5cm}total & 2.30 & 2.16 & 2.28 & 2.16\\[0.0ex]
$\mu_0$\\
\hspace*{0.5cm}random& 0.53 & 0.43 & 0.54 & 0.45\\
\hspace*{0.5cm}system& 0.24 & 0.23 & 0.25 & 0.23\\
\hspace*{0.5cm}cosmic (due to $x$)   & 0.32 & 0.30 & 0.32 & 0.30\\
\hspace*{0.5cm}cosmic (in $\mu$) & 2.15 & 2.05 & 2.13 & 2.04\\
\hspace*{0.5cm}total & 2.25 & 2.13 & 2.23 & 2.12\\[1.0ex]
FK6 = FK5+HIP        & 0.68 & 0.59 & 0.62 & 0.57\\[1.0ex]
ratio of HIPPARCOS& 1.2\hphantom{0} &
                    1.1\hphantom{0} & 1.1\hphantom{0} &
                    1.1\hphantom{0}\\
\hspace*{0.7cm}to FK6 errors\\[0.5ex]\hline
%\vspace*{-4.00 cm}
\end{tabular}
\end{table}
%%%%%%%%%%%%%%%%%%%%%%%%%%%%%%%% End of Table 3 %%%%%%%%%%%%%%%%%%%%%%%%%%%%%

%%%%%%%%%%%%%%%%%%%
% Begin of        %
%                 %
% T a b l e    4  %
%                 %
%%%%%%%%%%%%%%%%%%%
\begin{table}[t]
\caption{
Error budget for GC+HIP proper motions
}
\begin{tabular}{lcccc}
\hline\\[-1.0ex]
\multicolumn{5}{c}{Typical mean errors of proper motions}\\
\multicolumn{5}{c}{(in one component, averaged over $\mu_{\alpha\ast}$ and
$\mu_\delta$; units: mas/year)}\\[1.5ex]\hline\\[-2.0ex]
Sample of stars:& \multicolumn{2}{c}{29\,717 GC} & \multicolumn{2}
{c}{11\,773 GC}\\ [0.7ex]
\hline\\[-2.0ex]
& rms aver. & median & rms aver. & median\\[0.5ex]\hline\\[-1.0ex]
\multicolumn{5}{c}{\bf{SI}: \hspace*{0.3 cm} single-star mode}\\[0.0ex]
HIPPARCOS&       &      &      &     \\
\hspace*{0.5cm}random& 1.47  & 0.73 & 0.75 & 0.69\\[0.0ex]
GC\\
\hspace*{0.5cm}random& 10.57\hphantom{1} & 9.38 & 8.59 & 7.55\\
\hspace*{0.5cm}system& 0.57 & 0.49 & 0.56 & 0.48\\
\hspace*{0.5cm}total & 10.59\hphantom{1} & 9.39 & 8.61 & 7.57\\[0.0ex]
$\mu_0$\\
\hspace*{0.5cm}random& 1.78 & 1.74 & 1.42 & 1.44\\
\hspace*{0.5cm}system& 0.18 & 0.12 & 0.13 & 0.12\\
\hspace*{0.5cm}total & 1.79 & 1.75 & 1.43 & 1.45\\[1.0ex]
GC+HIP               & 0.72 & 0.63 & 0.62 & 0.58\\[1.0ex]
ratio of HIPPARCOS& 2.0\hphantom{0} &
                    1.2\hphantom{0} & 1.2\hphantom{0} &
                    1.2\hphantom{0}\\
\hspace*{0.7cm}to GC+HIP errors\\[0.5ex]\hline\\[-1.0ex]
\multicolumn{5}{c}{\bf{LTP}: \hspace*{0.3 cm} long-term prediction}\\[0.0ex]
HIPPARCOS&       &      &      &     \\
\hspace*{0.5cm}random& 1.47  & 0.73 & 0.75 & 0.69\\
\hspace*{0.5cm}cosmic (in $\mu$) & 1.75  & 1.60 & 1.80 & 1.66\\
\hspace*{0.5cm}total&  2.29  & 1.76 & 1.95 & 1.80\\[0.0ex]
GC\\
\hspace*{0.5cm}random& 10.57\hphantom{1} & 9.38 & 8.59 & 7.55\\
\hspace*{0.5cm}system& 0.57 & 0.49 & 0.56 & 0.48\\
\hspace*{0.5cm}total & 10.59\hphantom{1} & 9.39 & 8.61 & 7.57\\[0.0ex]
$\mu_0$\\
\hspace*{0.5cm}random& 1.78 & 1.74 & 1.42 & 1.44\\
\hspace*{0.5cm}system& 0.18 & 0.12 & 0.13 & 0.12\\
\hspace*{0.5cm}cosmic (due to $x$)   & 0.12 & 0.10 & 0.12 & 0.11\\
\hspace*{0.5cm}total & 1.79 & 1.75 & 1.43 & 1.45\\[1.0ex]
GC+HIP               & 1.17 & 1.15 & 1.06 & 1.05\\[1.0ex]
ratio of HIPPARCOS& 2.0\hphantom{0} &
                    1.5\hphantom{0} & 1.8\hphantom{0} &
                    1.7\hphantom{0}\\
\hspace*{0.7cm}to GC+HIP errors\\[0.5ex]\hline\\[-1.0ex]
\multicolumn{5}{c}{\bf{STP}: \hspace*{0.3 cm} short-term prediction}\\[0.0ex]
HIPPARCOS&       &      &      &     \\
\hspace*{0.5cm}random& 1.47  & 0.73 & 0.75 & 0.69\\[0.0ex]
GC\\
\hspace*{0.5cm}random& 10.57\hphantom{1} & 9.38 & 8.59 & 7.55\\
\hspace*{0.5cm}system& 0.57 & 0.49 & 0.56 & 0.48\\
\hspace*{0.5cm}cosmic (in $\mu$) & 1.75 & 1.60 & 1.80 & 1.66\\
\hspace*{0.5cm}total & 10.73\hphantom{1} & 9.53 & 8.79 & 7.75\\[0.0ex]
$\mu_0$\\
\hspace*{0.5cm}random& 1.78 & 1.74 & 1.42 & 1.44\\
\hspace*{0.5cm}system& 0.18 & 0.12 & 0.13 & 0.12\\
\hspace*{0.5cm}cosmic (due to $x$)   & 0.12 & 0.10 & 0.12 & 0.11\\
\hspace*{0.5cm}cosmic (in $\mu$) & 1.75 & 1.60 & 1.80 & 1.66\\
\hspace*{0.5cm}total & 2.51 & 2.37 & 2.30 & 2.20\\[1.0ex]
GC+HIP               & 0.82 & 0.69 & 0.69 & 0.65\\[1.0ex]
ratio of HIPPARCOS& 1.8\hphantom{0} &
                    1.1\hphantom{0} & 1.1\hphantom{0} &
                    1.1\hphantom{0}\\
\hspace*{0.7cm}to GC+HIP errors\\[0.5ex]\hline
%\vspace*{-4.00 cm}
\end{tabular}
\end{table}
%%%%%%%%%%%%%%%%%%%%%%%%%%%%%%%% End of Table 4 %%%%%%%%%%%%%%%%%%%%%%%%%%%%%

%\listofobjects

\end{document}